\documentstyle[aps,preprint]{revtex}
\begin{document}
%
\preprint{hep-th/9603128, CU-TP-742, SNUTP-96-24}
\draft
\title{\Large\bf The self-dual Chern-Simons $CP(N)$ models  }
\author{Kyoungtae Kimm$^1$, Kimyeong Lee$^2$,
               and Taejin Lee$^3$}
\address{
{$^1$ Department of Physics,
      Seoul National University, Seoul 151-741, Korea} \\
{$^2$ Physics Department,
      Columbia University, New York, New York. 10027, U.S.A.} \\
{$^3$ Department of Physics, Kangwon National University,
      Chuncheon 200-701, Korea} }
\maketitle
\begin{abstract}
We study the Chern-Simons $CP(N)$ models with a global $U(1)$ symmetry
and found the self-dual models among them.  The Bogomolnyi-type bound
in these self-dual models is a nontrivial generalization of that in
the pure $CP(N)$ models. Our models have quite a rich vacuum and
soliton structure and approach the many known gauged self-dual models
in some limit.
\end{abstract}


The $CP(N)$ models have many interesting structures.  In two
dimensional spacetime their action is conformally invariant and there
exist instantons which are topologically nontrivial\cite{zak,raja}.
They exhibit interesting phenomena common to Yang-Mills theories at
four dimensions, like confinement and asymptotic freedom\cite{witten}.
In three dimensions, the $CP(N)$ models have self-dual solitons whose
field configuration and interaction have been studied
well\cite{zak,ward}.

On the other hand, there have been considerable studies of the
relativistic self-dual gauged Higgs systems during the last few years.
The gauge group can be abelian or nonabelian
\cite{hong,klee,kao,dunne} with Maxwell or
Chern-Simons kinetic terms. The vacuum and soliton structure of the
models can exhibit a rich variety.  The solitons in these models carry
the fractional spin and satisfy the fractional statistics.

Recently some hybrids of the $CP(1)$ model, or the $O(3)$ sigma model
and abelian (Maxwell) Chern-Simons Higgs models as been
proposed\cite{sch,kimm,glad}.  These gauged self-dual $CP(1)$ models have
the vacuum and soliton structures which inherits the characteristics
of their parent models: There are skymion like solitons similar to
those in the $CP(1)$ models, topological and nontopological solitons
similar to those in the Chern-Simons Higgs models. In addition, there
are solitons in the broken phase, which cannot put into a rotationally
symmetric form\cite{kimm}.

In this paper, we generalize the gauged $CP(1)$ case to the gauged
$CP(N)$ models.  The gauge group can be abelian or nonabelian.  The
kinetic term for the gauge field is chosen to be the Chern-Simons term
for the convenience.  The crucial condition for the existence of
nontrivial structures beside what we get from the naive $CP(N)$ models
is the existence of at least one global $U(1)$ symmetry which commutes
with the gauge symmetry. This requirement of a global $U(1)$ symmetry
was also a key component in finding the nonabelian self-dual
Chern-Simons Higgs models\cite{klee}.  This global $U(1)$ charge could
be a part of a gauged abelian group.  This work contrasts the previous
works\cite{nardelli} on the gauged $CP(N)$ models, where there is no
global $U(1)$ symmetry and so the solitonic structure is identical to
that of the pure $CP(N)$ models.

The $CP(N)$ model concerns with the space of a $(N+1)$-dimensional
complex vector field $z= (z_1,z_2,...,z_{N+1})$ of unit length, with
the equivalence relation under the overall phase rotation, $z \sim
e^{i\alpha} z$. This complex projective space of complex dimension $N$
is equivalent to the coset space $SU(N+1)/U(1)\!\times\! SU(N)$.  ( It
is quite straightforward to generalize our consideration here to the
general Grassmanian models with the manifold $G(M,N) = U(N)/U(M)\times
U(N-M)$.)  There have been many studies of the $CP(N)$ models in three
dimensions\cite{zak,raja,witten,ward} whose Lagrangian is given as
\begin{eqnarray}
 {\cal L}_{CP(N)} = \nabla^\mu {\bar{z}} \nabla_\mu z ,
\label{eq:old}
\end{eqnarray}
where $\nabla_\mu z = \partial_\mu z - ({\bar{z}} \partial_\mu z) z$.  This
theory has a global $SU(N+1)$ symmetry and a local $U(1)$ symmetry
which removes the degrees of freedom corresponding to the overall
phase of $z$.  The self-dual finite energy configurations are given by
$z(\vec{x})=w/|w|$, where $w$ is a (anti)holomorphic rational function
of spatial coordinates $x+iy$.  In this theory, there is a conserved
topological current,
\begin{eqnarray}
 k^\mu = -i \epsilon^{\mu\nu\rho}\partial_\nu({\bar{z}}\partial_\rho z),
\label{eq:top1}
\end{eqnarray}
whose charge $S = \int d^2x k^0$ is called the degree and measures the
second homotopy class of the mapping $z(t,\vec{x})$ from the space
$R^2$ to the manifold $CP(N)$ when $z(t,\vec{x}=\infty)=z_0$ for a
constant vector $z_0$.

We generalize the Lagrangian (\ref{eq:old}) by adding the gauge
coupling, the Chern-Simons term and the potential energy.  The
Lagrangian we are to consider reads
\begin{eqnarray} 
{\cal L}= \frac{\kappa}{2}\epsilon^{\mu\nu\rho} \Bigl(A_\mu^a
\partial_\nu A_\rho^a  
+\frac{1}{3}f^{abc}A_\mu^a A_\nu^b A_\rho^c\Bigr)
+ |\nabla_\mu z|^2 -U(z),
\label{eq:lag}
\end{eqnarray} 
where the covariant derivative is defined as
\begin{eqnarray}
 \nabla_\mu z =D_\mu z -({\bar{z}} D_\mu z)z , 
\end{eqnarray} 
where $D_\mu z = \partial_\mu z -iA_\mu^a R^a z$.  The gauge group is
a proper subgroup of $SU(N+1)$, which is generated by the hermitian
traceless matrices $R^a$ satisfying the relation
$[R^a,R^b]=if^{abc}R^c$. We require that there is a conserved global
abelian symmetry whose generator is chosen to be $T^D\equiv
(1,1,\dots,1,-N)/\sqrt{2N(N+1)}$. If the gauge group is purely
nonabelian, the generators $R^a$ for the gauge group should have
vanishing elements on the $(N+1)$-th row and column so that the
$R^a$'s commute with $T^D$.  If there are more conserved global $U(1)$
symmetries, there are corresponding complicated models, which we not
pursued here. The potential $U(z)$ will be specified later by
requiring that there is a Bogomolnyi-type bound on the energy.

The kinetic energy for the gauge field can also have the Yang-Mills
term. In this case, we need also a neutral scalar field $N^a$ in the
adjoint representation of the gauge group for the self-duality. The
qualitative behavior does not change much from the pure Chern-Simons
case, and so we will not also pursue this direction in this paper.

The theory possesses the following conserved topological current
$K^\mu $
\begin{eqnarray} 
K^\mu=-i\epsilon^{\mu\nu\rho}\partial_\nu(\bar{z}D_\rho z),
\label{eq:top2}
\end{eqnarray} 
which is a gauge-invariant generalization of $k^\mu$ in Eq.~(\ref{eq:top1}).
The Gauss law constraint obtained from the 
variation of $A_0^a$ is
\begin{eqnarray}
  \kappa  F_{12}^a -i\left\{\nabla_0 {\bar{z}}
 \left(R^a z -z ({\bar{z}} R^a z)\right)
                      - {\rm h.c.}\right\}=0.
\label{eq:gauss1}
\end{eqnarray} 
Any physical field configurations should satisfy this constraint. This
constraint is preserved in the time evolution and so any physical
configuration at a given moment will remain to be physical.

There is also a conserved global $U(1)$ current for the generator
$T^D$, 
\begin{eqnarray}
J^\mu  =  i \left\{\nabla^\mu {\bar{z}}\left(T^D z  
                   -z({\bar{z}} T^D z)\right)
           -{\rm h.c.}\right\} .
\label{eq:zu1}
\end{eqnarray}
Note that $\nabla_\mu {\bar{z}} z = 0$, and so some terms in the Gauss law
(\ref{eq:gauss1}) and the global $U(1)$ current (\ref{eq:zu1}) vanish.
However, we will find later that keeping those terms is useful.

We start with the energy
functional
\begin{eqnarray}
E =\int d^2 x \left\{ 
  |\nabla_0 z|^2 +|\nabla_i z|^2 +U(z)\right\}. 
\end{eqnarray}
With the Gauss law, we get 
\begin{eqnarray}
|\nabla_i z|^2 &=& |(\nabla_1\pm i \nabla_2 )z|^2 \pm ( K_0 
   + F_{12}^a ({\bar{z}} R^a z) ) \nonumber \\
  &=& |(\nabla_1\pm i \nabla_2 )z|^2 
      \pm (K_0 + \frac{v}{\kappa}J_0 )
              \nonumber  \\
 && \pm \frac{i}{\kappa}\left\{ 
 \nabla_0{\bar{z}} \left[ (R^a z -z ({\bar{z}} R^a z))({\bar{z}} R^a z)
-v\left( T^D z -z({\bar{z}} T^D z)\right)\right]
 -{\rm h.c.}\right\} \label{eq:kin} ,
\end{eqnarray}
up to a total derivative which does not contribute to the energy.
Here we added and then subtracted a term proportional to the conserved
global $U(1)$ charge density. This allows us to introduce a free real
parameter $v$ to the model. (If there are more conserved global $U(1)$
current, we may introduce more free parameters like $v$.)

We choose the potential as 
\begin{eqnarray}
U(z)=\frac{1}{\kappa^2}\left|
 \left( R^a z - z ({\bar{z}} R^a z) \right) ({\bar{z}} R^az)
 -v\left(T^D z-z({\bar{z}} T^D z) \right) \right|^2 .
      \label{eq:pot}
\end{eqnarray}
This choice and Eq.~(\ref{eq:kin}) allow us to express the energy
density as
\begin{eqnarray}
{\cal E}  &=&  |(\nabla_1 \pm i \nabla_2) z|^2  \nonumber\\
 &&+ \left| 
 \nabla_0z \pm\frac{i}{\kappa}
 \left\{ (R^a z -z ({\bar{z}} R^a z))({\bar{z}} R^a z)
-v\left( T^D z -z({\bar{z}} T^D z)\right)\right\}
 \right|^2         \nonumber \\
&&  \pm (K_0+ \frac{v}{\kappa} J_0 ) .
\end{eqnarray}
Since the first two terms in the right hand side are nonnegative,
there is a Bogomolnyi-type energy bound $E\ge |T| $, where the
`topological charge'
\begin{eqnarray}
T=\int d^2x (K_0 +\frac{v}{\kappa}J_0)  
\end{eqnarray}
is a generalization of the degree $S$.

The field configurations saturating the energy bound satisfy the Gauss
law and the `self-dual' equations
\begin{eqnarray}
&&(\nabla_1 \pm i \nabla_2)z =0 \label{eq:zself1}  ,     \\ 
&& \nabla_0 z
\pm\frac{i}{\kappa}
 \left\{ (R^a  - {\bar{z}} R^a z)({\bar{z}} R^a z) -v\left(
T^D  -{\bar{z}} T^D z\right)\right\} z =0 . \label{eq:zself2}
\end{eqnarray}
The reason for keeping the vanishing terms in Eqs.~(\ref{eq:gauss1})
and (\ref{eq:zu1}) is obvious now.  From Eq.~(\ref{eq:zself2}) the
identity ${\bar{z}} \nabla_0 z = 0 $ does not lead any condition on the $z$,
because the rest of Eq.~(\ref{eq:zself2}) also vanishes when it is
multiplied by ${\bar{z}}$. This trick  does not appear in 
Ref.~\cite{nardelli}, resulting in a nontrivial condition on $z$. Combined
with the Gauss law, Eq.~(\ref{eq:zself2}) 
gives
\begin{eqnarray}
\kappa^2 F^a_{12}&=&
\mp \left\{ \left({\bar{z}}(R^a R^b+R^bR^a)z
 -2({\bar{z}} R^a z)({\bar{z}} R^b z)\right)({\bar{z}} R^b z)\right.
\nonumber \\
&&\left.\hspace{8mm} -v\left( {\bar{z}} (R^a T^D +T^D R^a )z
          -2({\bar{z}} T^D z)({\bar{z}} R^a z)\right)
 \right\} . \label{eq:zgauss}
\end{eqnarray}

If the gauge group were the full $SU(N+1)$, there would be no
conserved abelian global current and so $v=0$. One can show in this
case the field strength vanishes.  This makes the gauge field part of
the self-dual configurations trivial.  The potential of this case
turns out to be a constant. The soliton structure turns out to be
identical to that of the $CP(N)$ model\cite{nardelli}.

When Eq.~(\ref{eq:zself2}) is satisfied, the $U(1)$ charge density from
Eq.~(\ref{eq:zu1}) becomes
\begin{eqnarray}
\kappa J_0 &=&\mp 2\left\{ \left( {\bar{z}} (R^a T^D ) z
       -({\bar{z}} T^D z)({\bar{z}} R^a z) \right) ({\bar{z}} R^a z) \right. 
\nonumber
\\ 
 && \hspace{8mm} \left.
      -v\left( {\bar{z}} 
(T^D T^D) z -({\bar{z}} T^D z )( {\bar{z}} T^D z ) \right)
         \right\} .      
\label{eq:zcharge}
\end{eqnarray}
For the configuration satisfying
Eqs.~(\ref{eq:zself1}) and (\ref{eq:zself2}),
the total angular momentum becomes
\begin{eqnarray}
J&=&-\int d^2x \epsilon_{ij} x^j
 (\nabla_0 {\bar{z}} \nabla_j z+ \nabla_j {\bar{z}} \nabla_0 z)\nonumber \\
 &=& \frac{1}{2\kappa} \int d^2x x^i \partial_i
     \left( ({\bar{z}} R^a z)^2-2v ({\bar{z}} T^D z) \right) .
\label{eq:ang}
\end{eqnarray}

To see the $O(3)$ sigma model studied in Ref.~\cite{sch,kimm}, we
consider the $CP(1)$ model with a local abelian gauge symmetry with it
generator $T^D = {\rm diag}(1,-1)/2$. We can parameterize the
configuration space by a unit vector field $\vec{\psi}={\bar{z}}
\vec{\sigma} z$. With these identification we can construct the
following self-dual equations from Eqs.~(\ref{eq:zself1}) and
(\ref{eq:zself2});
\begin{eqnarray}
&& D_1 \vec\psi \pm \vec\psi \times D_2 \vec\psi =0  ,    \\ 
&& D_0 \vec\psi=\mp 
 \frac{1}{2\kappa}(2v-\vec n\cdot\vec \psi)(\vec n\times \vec\psi)  ,
\end{eqnarray}
with the Gauss law constraint
\begin{eqnarray}
2\kappa F_{12}+\vec n \cdot \vec\psi \times D_0\vec\psi =0 ,
\end{eqnarray}
where $D_\mu \vec\psi =\partial_\mu \vec\psi +A_\alpha \vec n
\times \vec\psi $ and $\vec n=(0,0,1)$.  With substituting
$v\rightarrow v/2$ and $\kappa\rightarrow\kappa/2 $, we can see that
these are the same self-dual equations 
as the ones considered in Ref.~\cite{kimm}.

To understand the general implications of the theory when the gauge
group is purely nonabelian, we reparameterize the $N+1$ dimensional
vector as $z= e^{i\alpha}(\phi, \sqrt{1-|\phi|^2})$ where $\phi$ is a
$N$-dimensional complex vector such that $|\phi|\le 1$.  As argued
before, the gauge generators $R^a$ have the vanishing $N+1$-th column
and row, and so the we can define the covariant derivative of $\phi$
as $D_\mu\phi = \partial_\mu \phi - i A^a_\mu R^a \phi$. The overall
phase $\alpha$ has been gauged way. The self-dual equations
(\ref{eq:zself1}) and (\ref{eq:zself2}) for the first $N$ components
become
\begin{eqnarray}
&&(D_1\pm i D_2 )\phi =
    \frac{1}{2}\left\{ {\phi^\dagger}(D_1\phi \pm i D_2\phi)
   -(D_1{\phi^\dagger}\pm i D_2 {\phi^\dagger})\phi\right\} \phi  ,
   \label{eq:sunself1}\\
&& D_0 \phi -\frac{1}{2}({\phi^\dagger} D_0\phi-D_0{\phi^\dagger} \phi)\phi
\pm\frac{i}{\kappa}
       \left\{(R^a-({\phi^\dagger} R^a \phi))
                   ({\phi^\dagger} R^a \phi)
        -v \textstyle{ \sqrt{\frac{N+1}{2N}} }
        (1-|\phi|^2) \right\}\phi=0  .
                 \label{eq:sunself2}
\end{eqnarray}
The $(N+1)$-th component of the self-dual equations follow from the
above equations.  With this representation of $z$, the Gauss law
(\ref{eq:zgauss}) and the charge density (\ref{eq:zcharge}) also get
somewhat simplified.

The potential energy (\ref{eq:pot}) in the pure nonabelian case can be
expanded and put together as
\begin{eqnarray}
U=&&\frac{1}{\kappa^2}\left|
         \left( R^a \phi -\phi({\phi^\dagger} R^a \phi)\right) 
        ({\phi^\dagger} R^a \phi) -v\textstyle{\sqrt{\frac{N+1}{2N}} }
          \phi  \left(1- |\phi|^2 \right)\right|^2 \nonumber \\
&& + \frac{1-|\phi|^2}{\kappa^2}\left|
      ({\phi^\dagger} R^a \phi)({\phi^\dagger} R^a \phi) 
       - v \textstyle{\sqrt{\frac{N+1}{2N}}} |\phi|^2 \right|^2 
\label{eq:pot2}.
\end{eqnarray}
Since the potential is nonnegative, the minimum of the potential could
be zero. The ground configuration with the zero potential energy will
satisfies the $(N+1)$ conditions,
\begin{eqnarray}
&& \left( R^a \phi -\phi({\phi^\dagger} R^a \phi)\right) 
    ({\phi^\dagger} R^a \phi) -v\textstyle{\sqrt{\frac{N+1}{2N}} }
     \phi  \left(1- |\phi|^2 \right)=0  , \label{eq:v1}\\
&&\sqrt{1-|\phi|^2} \left\{({\phi^\dagger} R^a \phi)({\phi^\dagger} R^a \phi) 
    - v \textstyle{\sqrt{\frac{N+1}{2N}}} |\phi|^2      
       \right\}=0 ,\label{eq:v2}
\end{eqnarray}
not all of which are independent of each other. (Here $\phi$ denotes
the vacuum expectation value of the $\phi$ field.)  One can easily show
that $F^a_{12}$ and $J^0$ for the ground configurations vanish as
expected.  When $|\phi|\neq 1$, we can substitute the second equation
to the first to get a simplified vacuum equation on the expectation
value,
\begin{eqnarray} 
 R^a \phi ({\phi^\dagger} R^a \phi) - v \textstyle{\sqrt{\frac{N+1}{2N}} }
     \phi =0        
\label{eq:vac1}      ,
\end{eqnarray}
which implies Eqs. (\ref{eq:v1}) and (\ref{eq:v2}).   When
$|\phi|=1$, instead we get  a different condition 
\begin{eqnarray}
 \left( R^a \phi -\phi({\phi^\dagger} R^a \phi)\right) 
    ({\phi^\dagger} R^a \phi)=0    \label{eq:vac2}       .
\end{eqnarray}
These equations (\ref{eq:vac1}) and (\ref{eq:vac2}) determine the
vacuum structure of the model.

To understand the vacuum and soliton structure of the theory, let us
consider the  Higgs limit, $0<v \ll 1$ and $|\phi| \ll 1$.
The kinetic part of the Lagrangian (\ref{eq:lag}) becomes $\nabla_\mu {\bar{z}}
\nabla^\mu z = D_\mu \phi^\dagger D^\mu \phi  + O(\phi^4) $.  From the
potential energy (\ref{eq:pot2}) we see that the small $\phi, v$ limit
is consistent if $\phi \sim O(\sqrt{v})$ and that the potential energy
in this limit becomes
\begin{eqnarray} 
U =  \frac{1}{\kappa^2} \left| R^a \phi ({\phi^\dagger} R^a \phi) - v
\textstyle{\sqrt{\frac{N+1}{2N}} }  \phi \right|^2
, \label{eq:pot3}
\end{eqnarray}
to order $v^4$. This is exactly the potential appeared in the
self-dual Chern-Simons Higgs models \cite{klee}.  Thus in this limit,
our models become self-dual Chern-Simons Higgs systems. These models
have been studied before and shown to have the rich vacuum and
solitonic structures \cite{klee,kao,dunne}.  As the vacuum condition
(\ref{eq:vac1}) is identical to that in the the Higgs case, the vacuum
structure of the Higgs systems will survive in our models. Thus if
$|\phi|<1$ is satisfied at the vacuum, the vacuum structure of the
Higgs limit is identical to that of our model. Especially a rich
vacuum structure appears when the matter field is in the adjoint
representation of the gauge group \cite{kao,dunne}.

Having see that the vacuum structure of the case $|\phi|<1$, we ask
whether there is any nontrivial vacuum structure in the case
$|\phi|=1$. We have analyzed Eq.~(\ref{eq:vac2}) in detail in the
$CP(3)$ case with the $\phi$ field in the adjoint representation of
$SU(2)$. It turns out that Eq.~(\ref{eq:vac2}) has two class of
solutions: one which is a continuation of the solutions of
Eq.~(\ref{eq:vac1}), and another of a different characteristic.  It
would be interesting to understand the solutions of
Eq.~(\ref{eq:vac2}) in more general situation.

To get another perspective of the models proposed here, we have
considered various choices of the gauge group, studied some
characteristics of those models, and list them as follows: (I) The
simplest case occurs when the theory is not gauged at all. Since
Eqs.~(\ref{eq:kin}) make sense even if we put $R^a=0$, our energy
bound and self-dual equations still work.  Our models in this case can
be regarded as the generalization of the self-dual nongauged $O(3)$
model studied by Leese~\cite{leese}. (II) We can gauge with $T^D$ as
its generator.  Since the vacuum manifold in the broken phase can be
shown to be simply connected for $CP(N)$ with $N\ge 2$, this theory
would lead to the semilocal strings in the broken phase\cite{tanmay}.
(III) We can gauge the $SU(N)$ subgroup under which the first $N$
components of the complex vector transform as a fundamental
representation. There is a global $U(1)$ symmetry generated by $T^D$.
(IV) We can also gauge this global $U(1)$ symmetry, ending up
with a gauge group $SU(N)\times U(1)$. (V) For $CP(N^2-1)$, we can
gauge $SU(N)$ with the first $N^2-1$ components belongs to the adjoint
representation of the gauge group. This case has a rich vacuum and
soliton structure.

In short, we have constructed the self-dual gauged Chern-Simons
$CP(N)$ models. The crucial ingredient is that there is at least one
global $U(1)$ charge which commutes with the local gauge symmetry.
These models generalize nontrivially both the $CP(N)$ models and the
self-dual Chern-Simons Higgs models. The vacuum and soliton structure
of the models can be quite rich, depending on the parameter $v$ and
the gauge symmetry. We have seen that many known self-dual
Chern-Simons Higgs models can be obtained by taking a suitable limit.

There are many directions to explore from this point.  There may be
some new vacuum structures which are not apparent in the parent
models.  Naturally we expect all self-dual solitons of the parent
models to be present in our model and there may be also a new type of
solitons as in the $CP(1)$ models. We hope to study these in some
concrete model.  The models we may obtain from the dimensional
reduction to two dimensional space time will be interesting as they
are renormalizable.  Classically these dimensionally reduced models
will have rich domain-wall solitons.

\acknowledgements{KK was supported in part by KOSEF through the Center for
Theoretical Physics of Seoul National University. KL was supported in part by
the NSF Presidential Young Investigator Program and the Alfred P. Sloan
Foundation. TL was supported by the Basic Science Research Institute Program,
Ministry of Education, 1995, Project BSRI-95-2401.}

\end{document}